# Fast stimulated Raman imaging for intraoperative gastro-intestinal cancer detection

Barbara Sarri[1], Rafaël Canonge[1], Xavier Audier[1], Emma Simon[1], Julien Wojak[1], Fabrice Caillol[2], Cécile Cador[2], Didier Marguet[3], Flora Poizat[2], Marc Giovannini[2,*] and Hervé Rigneault[1,*]

[1]Aix Marseille Univ, CNRS, Centrale Marseille, Institut Fresnel, Marseille, France

[2]Institut Paoli-Calmettes, Endoscopy and Gastroenterology Departement, Marseille, France

[3]Aix-Marseille Université, INSERM, CNRS, Centre d'Immunologie de Marseille-Luminy, Marseille, France

*Corresponding authors: giovanninim@ipc.unicancer.fr, herve.rigneaul@fresnel.fr

**Abstract:**

Conventional haematoxylin, eosin and saffron (HES) histopathology, currently the 'gold-standard' for pathological diagnosis of cancer, requires extensive sample preparations that are achieved within time scales that are not compatible with intra-operative situations where quick decisions must be taken. Providing to pathologists a close to real-time technology revealing tissue structures at the cellular level with HES histologic quality would provide an invaluable tool for surgery guidance with evident clinical benefit. Here, we specifically develop a stimulated Raman imaging based framework that demonstrates gastro-intestinal (GI) cancer detection of unprocessed human surgical specimens. The generated stimulated Raman histology (SRH) images combine chemical and collagen information to mimic conventional HES histopathology staining. We report excellent agreements between SRH and HES images acquire on the same patients for healthy, pre-cancerous and cancerous colon and pancreas tissue sections. We also develop a novel fast SRH imaging modality that captures at the pixel level all the information necessary to provide instantaneous SRH images. These developments pave the way for instantaneous label free GI histology in an intra-operative context.

**Introduction:**

Most organs from the gastrointestinal tract are subjected to cancer development. Nowadays, colorectal cancer is the fourth cause of cancer related-death worldwide and pancreas cancers are fatal with a survival rate of only 1-5% after 5 years[1, 2, 3]. Conventional histopathology is the current 'gold-standard' for pathological diagnosis of cancer, it requires the removal of small regions of suspect tissues (biopsies) that are later sectioned and stained with haematoxylin, eosin and saffron (HES)[4] to provide a pathologist with a detailed tissue cellular morphology view from which the diagnosis is made. Larger tissues or entire group of gastric organs can also be removed during surgery and post-operative HES is performed to assess the evolution of the tumour, its margin and assign post operation treatments (like chemotherapy). The gold standard HES protocol requires the excised samples to undergo several preparation steps (dehydration, fixation, wax, staining, slicing, mounting) that are labour and time consuming, typically in the range of 10 hours up to 3 days[4]. A faster histology protocol, known as extemporaneous histology (EH), can be performed in ~30 mm and is used in an intra-operative context to guide surgeries. However EH uses a single stain (toluidine blue[5]) that is prone to artefacts and provides less information that HES protocols.

There is a need for fast diagnostic techniques that would deliver quick and accurate 'histology like' images to assist in surgical and decision making. In the GI context rapid access to histologic data would allow for the detection of peritoneal metastasis during gastric surgeries that is a critical point to stop the surgery and orient the patient toward chemotherapy treatments. Histopathologic diagnostic requires the visualization of cell nuclei and cell bodies in a tissue section. Very recently an elegant approach to identify these components in a label free manner and at high resolution has been proposed using stimulated Raman scattering (SRS)[6,7]. SRS[8,9] can probe the distinctive chemical bonds found in cell cytoplasm and cell nuclei providing a unique label free molecular contrast to identify tissues macroscopic and microscopic architecture. Most importantly the unique ability of SRS to image $CH_2$ and $CH_3$ chemical bonds found predominantly in cell bodies and cell nuclei, respectively, was shown to provide rapid label free microscopic images of healthy and cancer brain sections in near-perfect concordance with standard histology[7,10]. This major step forward opens the completely new field of stimulated Raman histology (SRH) for real-time label-free intra-operative surgical cancer tissue resections and treatments[11].

The SRS technology has never been developed to image GI cancer tissues. Contrary to brain imaging GI tissues are more diverse in shape and contain collagen. Collagen is clearly seen in conventional HES using the saffron stain and its macroscopic and microscopic organization is an important indicator to reveal cancer development in GI tissue such as the pancreas.

In this paper we developed for the first time a SRH framework combining SRS and second harmonic generation (SHG) to generate images of colon and pancreas tissue frozen and fresh sections with near perfect agreement with conventional HES based techniques. We also report novel SRS implementations, λ-switch SRS and frequency modulated SRS (FM-SRS) that enable to generate SRH images over a millimetre field of view and in a time scale compatible with intraoperative workflow. The new developed SRS implementations allow for the acquisition of SRS images at the two wavenumbers corresponding to $CH_2$ and $CH_3$ chemical bonds, simultaneously. For each samples, toluidine blue and HES protocols were performed in parallel for comparison and assessment with SRH. Our work demonstrates an excellent agreement between SRH and HES for both normal and cancer tissues. It also paves the way to use SRH for GI tissues cancer identification as a possible alternative to extemporaneous histology to provide accurate HES like images to assist in surgical decision making, in an intra-operative context.

**Materials & Methods:**

*Cryogenic and fresh samples:* Fresh colon and pancreas tissues were excised from the patients after surgery and solidified in liquid nitrogen. The cryogenics tissues were carefully stick onto a mount, sliced down to ~15µm thick sections using a microtome and mounted onto microscope slides without further preparation. For each sample imaged with SRS and SHG two adjacent slides were stained using toluidine blue and regular HES for comparison with SRH. Toluidine blue and HES images were analysed using the Calopix software. Fresh tissues pieces (~1mm size) excised from surgery were directly mounted on a microscope slides and squeezed under a coverslip to reach a thickness of ~100µm. All tissue sections were collected in the context of Institut Paoli-Calmettes approved protocol from patients who provided informed consent. Tissues in excess of what was needed for diagnosis were eligible for imaging. All methods were performed in accordance with the EU guidelines and regulations

COMMISSION DIRECTIVE 2006/17/EC of 8 February 2006 as regards technical requirements for the donation, procurement and testing of human tissues and cells.

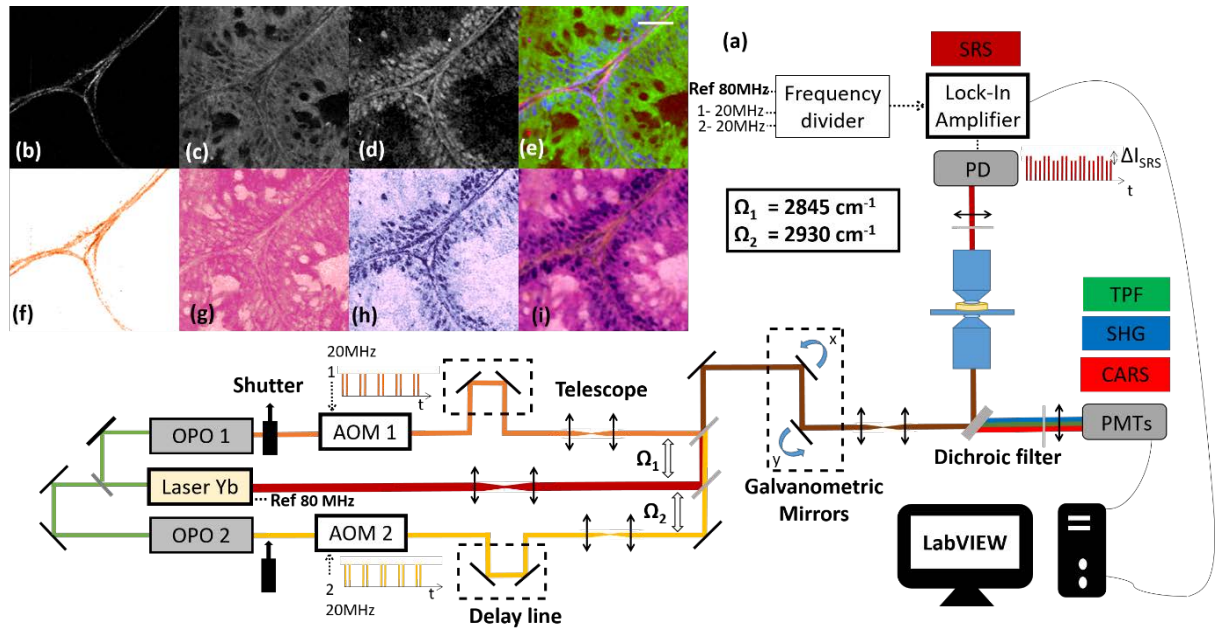

**Figure 1:** Performing stimulated Raman histology (SRH) on cryogenic and fresh samples. (a) Schematic diagram of the set-up allowing to probe simultaneously the CH$_2$ ($\Omega_1$=2845cm$^{-1}$) and CH$_3$ ($\Omega_2$=2930cm$^{-1}$) chemical bonds: OPO: optical parametric oscillator, AOM: acousto-optic modulator, PD: photodiode, TPEF: two-photon excited fluorescence, SHG: second harmonic generation, CARS: anti-Stokes Raman scattering, SRS: Stimulated Raman Scattering. Built-in mechanical shutters in OPO$_1$ and OPO$_2$ allow to perform the λ-switch modality, 180° phase shift between AOM$_1$ and AOM$_2$ enables the FM-SRS modality. (b-e) Data acquisition on a human colon tissue: (b) SHG signal reveals the collagen, (c) SRS signal at $\Omega_1$=2845cm$^{-1}$ (SRS$_{2845cm^{-1}}$) shows the cell bodies, (d) subtracting the SRS signals at $\Omega_2$=2930cm$^{-1}$ and $\Omega_1$=2845cm$^{-1}$ allows to highlight the nuclei distribution. (e) Composite image built from (b-d) SHG (magenta), SRS$_{2845cm^{-1}}$ (green), SRS$_{2930cm^{-1}}$ – SRS$_{2845cm^{-1}}$ (blue). (f-i) HES virtual coloring: Applying specific look up tables (LUTs) to (b,c,d) mimics HES staining: orange (SHG) to mimic the saffron, pink (SRS$_{2845cm^{-1}}$) to mimic the eosin, dark purple (SRS$_{2930cm^{-1}}$ – SRS$_{2845cm^{-1}}$) to mimic hematoxylin allow to build a label-free SRH (Stimulated Raman Histology) histological-like image (i) that can be readily interpreted by the histo-pathologist. Scale bar is 30 µm.

***Nonlinear label free imaging:*** Stimulated Raman scattering (SRS), second harmonic generation (SHG), coherent anti-Stokes Raman scattering (CARS) and two-photon excited fluorescence (TPEF) imaging were performed on a custom-built 3 colors set-up derived from [12]. As shown in figure 1-a, a mode-locked ytterbium (Yb) fiber laser (Emerald Engine, APE) operating at 1031 nm was, on the one hand, used as one of the colour and, on the other hand, was frequency doubled to pump two tuneable optical parametric oscillators (OPO$_1$ and OPO$_2$, Emerald, APE). OPO$_1$ and OPO$_2$ provide two distinct colours and act as the pump beams whereas the Yb laser provides the Stokes beam (1031 nm). All beams consist of picosecond pulse trains (pulse duration 2 ps, repetition rate 80 MHz). The beam from OPO$_1$ was set at 797.3 nm so that the pair OPO$_1$/1031 nm probed the lipid vibrational mode (CH$_2$) at 2845 cm$^{-1}$, while the beam from OPO$_2$ was set at 792.2 nm so that the pair OPO$_2$/1031 nm probed the CH$_3$ protein vibrational mode at 2930 cm$^{-1}$. The three beams were recombined using dichroic mirrors and brought into the microscope (Nikon, inverted TiU). In order to recombine the beams from the two OPOs, whose wavelengths were very close, a band-pass dichroic filter (TBP01-900/11, Semrock), whose 20 nm transmission windows was angle dependant, was used. Both OPO beams were modulated at 20MHz by acousto-optics modulators (AOM) (AA Optoelectronic MT200-AO). Delay lines were placed after the AOMs to ensure temporal overlap with the 1031 nm Stokes beam. The AOM modulators were driven directly from the laser repetition rate frequency (80 MHz) using a f/4 frequency divider (APE, 137761). The two 20MHz outputs of the frequency divider were set with a relative phase shift of 180° such that AOM$_1$ and AOM$_2$ modulations are 180° phase shifted, this features is important for the FM-SRS modality. The stimulated Raman gain (SRG) signals (OPO$_1$/1031 nm and OPO$_2$/1031 nm) were detected on the Stokes beam using a high speed photodiode and a fast lock-in amplifier (APE, LIA). The SRS signals were detected in the forward direction while TPEF, CARS and SHG signals were detected in the epi-direction. Excitation and Epi collection were performed using a numerical aperture (NA)=1.15 objective lens (Nikon APO LWD water 40x). Dichroics and bandpass filters (FES 700, LP420, FF593 , Thorlabs) allowed to differentiate between the TPEF, SHG and CARS signals that were recorded using photomultiplier tubes (PMTs) (Hamamatsu H7421-40 and H7421-50). In the forward direction a condenser lens NA=1.10 (Nikon APO LWD, water 60x) collected the light. A set of filters (SP 980, LP 860, Semrock) allowed to filter the 1031 nm Stokes beam before the detection photodiode. The beams were raster scanned on the samples using a set of galvo mirrors (Cambridge Technology 6215HM60). In these conditions, the SRS, CARS, TPEF and SHG spatial resolution was about 0.4 µm along the transverse direction (x and y axis) and 5 µm in the longitudinal direction (z axis) over a field of view (FoV) of 100 µm x 100 µm. Powers from each beam at the sample plane were set between 20 mW and 25 mW. A custom made Labview program[12, 13] controlled the galvo scanner, the excitation objective z positioning (Pifoc, PI), the OPOs and the microscope motorized stage (Märzhäuser, Tango2). The later allowing millimeter-scale images by stitching 100 µm x 100 µm images next to each other.

Two modalities were used to image tissue sections. The first one, named λ-switch (Fig. S1 (1)), uses the OPOs internal mechanical shutters to active sequentially OPO$_1$/1031 nm and OPO$_2$/1031 nm for each FoV. With this the SRS$_{2845cm-1}$ (CH$_2$) and the SRS$_{2930cm-1}$ (CH$_3$) signals are acquired sequentially together with the SHG signal and post-processed to generate SRH images mimicking HES stains. In this modality the total (SRS$_{2845cm-1}$, SRS$_{2930cm-1}$, SHG) imaging time for a 100 µm x 100 µm (200 x 200 pixels per 200 pixels) FoV was 20 s (dwell time per pixel 40 µs, 3 accumulations). A 1 mm x 1 mm stitched image could be performed in 50 min. The OPO optical powers prior to λ-switch acquisitions were measured on a daily basis by recording the SRS signal on a test sample (oil) at the same wavenumber (2930 cm$^{-1}$).

The second modality, named FM-SRS (Fig. S1 (2)), allows for the direct acquisition of the SRS$_{2845cm-1}$ and the SRS$_{2930cm-1}$ signal difference. This is done by setting a 180° phase shift between AOM$_1$ and AOM$_2$, in this case the lock-in amplifier directly provides the SRS signal difference SRS$_{2930cm-1}$-SRS$_{2845cm-1}$ allowing the direct identification of cell nuclei, a key feature to diagnose cancer tissue proliferation. Combined with SHG and TPEF (or CARS), FM-SRS allows to generate SRH images with HES quality instantaneously, at the pixel level, without the need of sequential acquisitions. In this modality the total (SRS, SHG, CARS, TPEF) imaging time for a 100 µm x 100 µm (200 x 200 pixels) FoV was 9 s (dwell time per pixel 40 µs, 3 accumulations) enabling a 1 mm x 1 mm stitched image acquisition is 25 minutes. Supplementary Fig. S1 shows the λ-switch and FM-SRS implemented modalities to generate SRH images.

***Data processing and virtual HES colouring:*** MatLab, ImageJ and Photoshop softwares were used to process the data. Firstly a dedicated MatLab code performing filtering into the image Fourier domain was used to remove the grid appearing between the 100 µm x 100 µm FoVs when stitched together to display millimetres large tissue images. In the λ-switch modality, the SHG (collagen - Fig. 1-b), SRS$_{2845cm-1}$ (CH$_2$ bonds - Fig. 1-c) and SRS$_{2930cm-1}$ (CH$_3$ bonds) are the raw images directly acquired by the microscope from which the SRS$_{nuclei}$= SRS$_{2930cm-1}$ – SRS$_{2845cm-1}$ is generated to reveal the cell nuclei (Fig. 1-d). These images are overlaid in Fig. 1-e (red: SHG, blue: SRS$_{nuclei}$, green: SRS$_{2845cm-1}$). Secondly, homemade look up tables (LUTs) were applied to both the SHG, SRS$_{2845cm-1}$ and SRS$_{nuclei}$ to mimic HES staining using ImageJ. A LUT in the orange/brown shades was thus allocated to the SHG data to imitate the saffron staining of collagen (Fig. 1-f). Meanwhile a LUT in the pink shades was assigned to the SRS$_{2845cm-1}$ image to resemble the eosin staining of both cytoplasm and cellular body (Fig. 1-g). Finally a LUT in the dark purple shades was applied to the SRS$_{nuclei}$ to mirror the hematoxylin staining of cell nuclei (Fig. 1-h). Images were then transferred into the Photoshop software to take advantage of the layers overlay capability. A threshold was applied to each layer to retrieve the relevant signal from the background noise. In order to differentiate the collagen fibres from the cell nuclei (which both appear in SRS$_{nuclei}$ – Fig. 1-d) priority was granted to the SHG image and images were overlaid (from top to bottom) in the following order: SHG, SRS$_{nuclei}$ and SRS$_{2845cm-1}$. The composite image was then transferred back to ImageJ for final rendering (smoothing and colour adjustments) to obtain the final SRH image (Fig. 1-i). In the specific case of pancreas sample that are rich in collagen a preliminary step consisting in applying a mask to bring to zero the pixel in which SHG signal was present, was performed before

LUT application onto the $SRS_{nuclei}$ data. Supplementary Fig. S2 presents the virtual HES colouring process from the raw images (case of Fig. 2-e).

The FM-SRS modality data processing and virtual HES colouring follows similar steps. FM-SRS provides directly the $SRS_{nuclei}$ image that was coloured with the dark purple shades LUT to mirror hematoxylin staining of cell nuclei. SHG colouring was treated as before to reveal collagen. To highlight both cytoplasm and cellular body TPEF or CARS images were used and coloured with the pink shades look up table to resemble the eosin staining. Choosing between TPEF and CARS signals gives some flexibility to obtain the best image depending on the sample.

**Results:**

***COLON: SRH enables to diagnose normal, low grade dysplasia and adenocarcinoma tissues.***

Millimeter-scale areas of colon tissues from patients suffering from different pathologies were explored in SRH using the $\lambda$-switch modality. For each studied sample HES and toluidine blue stained data from the same centimetre-scale region were also performed and weighed against SRH data. Tissues were examined by trained histo-pathologists at both microscopic and macroscopic scales to perform accurate medical diagnosis. Figure 2 summarizes our results. For every sample SRH images are found in good accordance with HES results. In the SRH figure 2-a, the 'daisy-field' characteristic of healthy (i.e. non-cancerous) colon mucosa is clearly identified: the global architecture of the tissue is regular and no stromal reaction is evidenced. A magnification (Fig. 2-b) shows uniform and homogeneous crypts with diameters around 100µm in diameter (blue arrows (Fig. 2-a). Nuclei are regularly spaced around the glands and in basal position (toward the outside of the gland – grey arrows (Fig. 2-b)). The mucus secretion is also not affected, the vacuoles being uniformly present within the crypts (red arrows (Fig. 2-b)). All these features are in good agreement with HES images (Fig. 2 c-d) and reveal that SRH images of healthy colon tissue can be accurately interpreted by trained pathologists and directly connected with features seen on standard HES images. Figures S3 shows the SRH and HES images of a larger tissue section and highlights further the tissue heathy microscopic features on selected regions of interest (ROI).

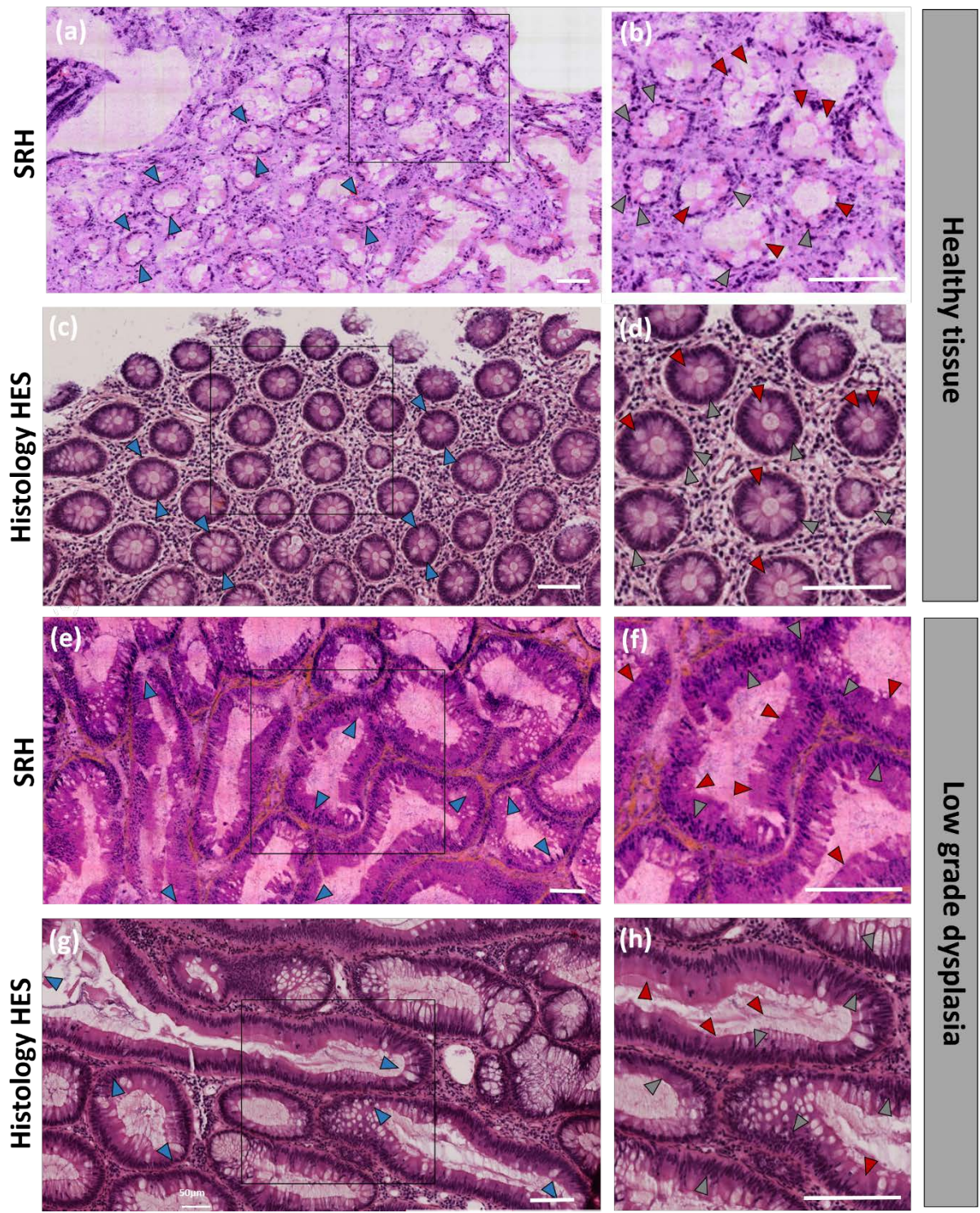

**Figure 2**: Comparing stimulated Raman histology (SRH) with haematoxylin, eosin and saffron (HES) on human colon tissue. (a) SRH image of a millimeter-scale area of healthy tissue, the daisy field is well defined. (b) Zoomed image on the region of interest (ROI) defined in (a). (c-d) HES images from the same region as (a-b) (within of a few mm³). Both SRH and HES images reveal key features of normal colon tissue such as homogeneous crypts diameters (blue arrows), regularly spaced nuclei around the glands (grey arrows), vacuoles uniformly present within the crypts (red arrows). (e) SRH image of a millimeter-scale area showing low grade dysplasia. (f) Zoomed image on ROI defined in (e). (g-h) HES images from the same region as (e-

f) (within of a few mm³). Both SRH and HES images reveal key features of low grade dysplasia tissue such as larger gland with irregular and inhomogeneous shapes (blue arrows), nuclear pseudo stratification (grey arrows) and reduction of the vacuoles within the glands (red arrows). Scale bar is 100 µm.

In Fig. 2-e, SRH image of a dysplasia colon tissue is exhibited while an HES image from the same patient is shown in Fig 2-g. Zoom in on selected ROI can be seen in Fig. 2-f and Fig. 2-h for SRH and HES, respectively. Contrary to the healthy case, the general tissue architecture is altered and the daisy field cannot be seen in the mucosa anymore. Lieberkühn glands are larger (blue arrows-Fig. 2 e-g), their shapes are both irregular and inhomogeneous in size which indicates a dysplasia. Looking closely at the nuclei distribution is essential to differentiate between low and high grade dysplasia. Both Fig. 2-f (SRH) and Fig. 2-h (HES) reveal numerous and highly packed nuclei (as compared to one nucleus per cell regularly spaced in normal tissue – Fig. 2-b and 2-d) . Pseudo-stratifications (lines of nuclei) are also very noticeable in SRH and HES (grey arrows - Fig. 2 f-h) while the nuclei sizes and shapes are quite regular, all this leads to a low grade dysplasia diagnostic[14]. This is confirmed with the fact that no high polarity loss can be noticed proving that high grade dysplasia has not been reach. Additionally, muco-secretion loss identified with a reduction of the presence of vacuoles within the glands (red arrows - Fig. 2 f-h), attests for atrophy of the crypts lumina and also corroborates for pre-cancerous state. Complementary magnified areas from Fig. 2-e and Fig. 2-g are presented in supplementary Fig. S4. We conclude that SRH images provide all the relevant HES histology information at the macroscopic and microscopic tissue scale to perform dysplasia diagnostics.

To ensure that no artefact due to the virtual colouring data processing was introduced, SRS composite images were set against adjacent cryogenic slides stained with toluidine blue (Fig. S5). Similar architecture are observed for nuclei and cell bodies in terms of shape, size and distribution for all heathy (Fig. S5 a-b), cancerous (Fig. S5 c-d), and pre-cancerous (Fig. S5 e-f), tissues.

We concentrate now on a cancerous colon tissue. Figure 3 presents SRH (Fig. 3 a-c) and HES (Fig. 3 b-d) images of tissue sections where the general architecture is immediately seen as disorganized as compared to healthy tissue (Fig. 2-a). This authenticates for an infiltrative tumor originating from the epithelial Lieberkühn glands. The strong stroma reaction is here evidenced by the fibrosis around the tumorous glands (red arrows on Fig. 3 a -b). The lumen has disappeared from the glands whose aspects are more angulated and where nuclei have invaded most of the cytoplasmic cell volume (increase of the nucleocytoplasmic ratio). These are signs of malignancy and allow to ascertain for a colon adenocarcinoma diagnostic. Besides, tumorous necrotic cells, which attest of a lack of blood supply, are also visible both in SRH and HES (grey arrows on Fig 3 a-b). A closer examination allows to reinforce the previous conclusions (Fig. 3 c-d) and further indicates the presence of numerous nucleoli (chromatin condensation – blue arrows) within the glands. Nucleoli are decisive criteria for malignancy[15]. These results demonstrate further the excellent consistency between SRH and HES images to identify colon adenocarcinomas.

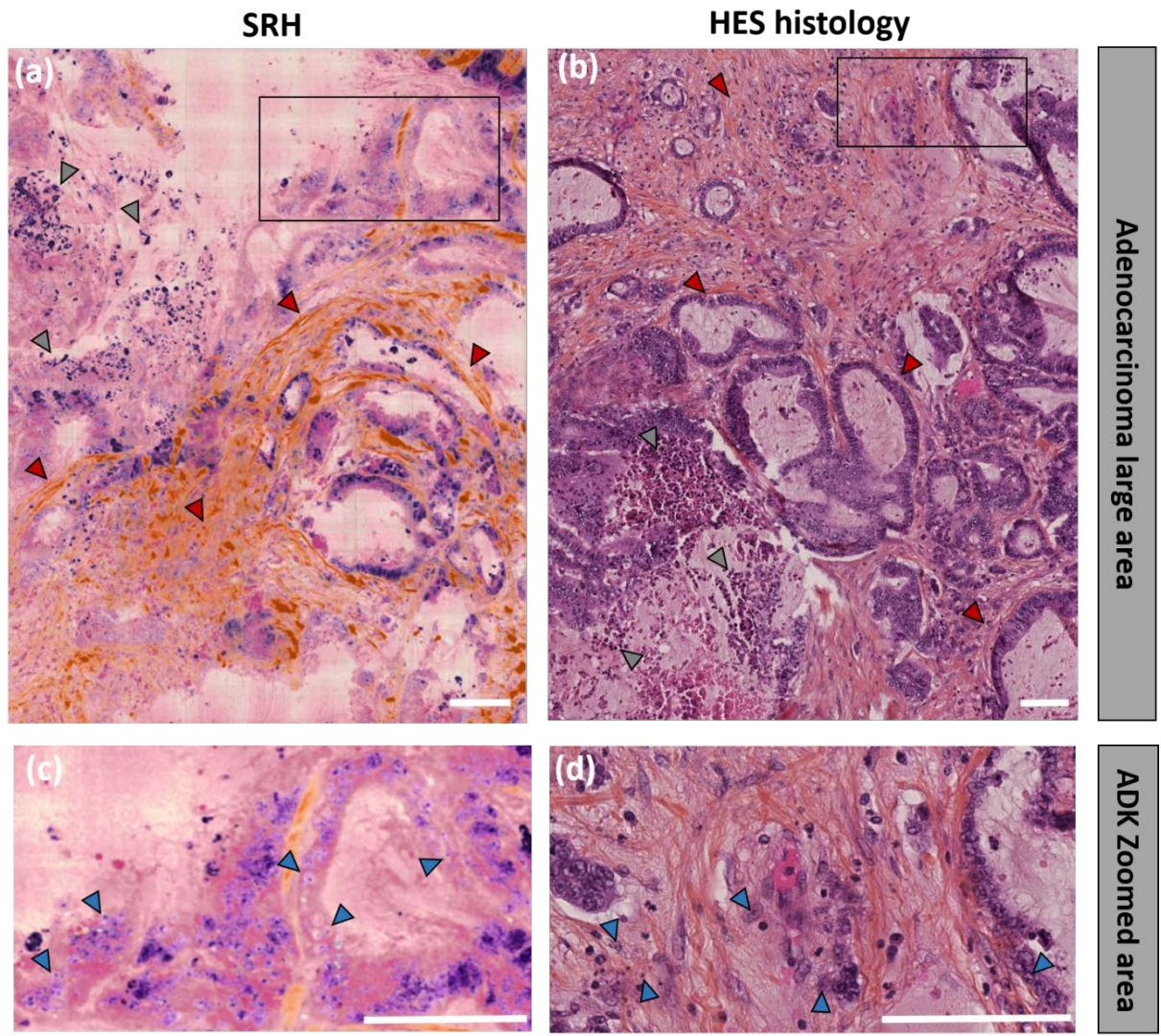

**Figure 3:** Human colon adenocarcinoma imaging. (a) SRH and (b) HES images from the same region (within of a few mm³). On both SRH and HES images signs of malignancy are present: a strong stroma-fibrosis around the glands (red arrows) as well as necrotic tumor cells (grey arrows) can be visualized. (c) and (d) are zoomed on the ROIs defined in (a) and (b), respectively in which large number of nucleoli can be evidenced (blue arrows). Scale bar 100 μm.

*PANCREAS: SRH enables to diagnose normal and adenocarcinoma tissues*

Figure 4-a present a SRH image of a healthy pancreas, its architecture correlates well with the HES image (Fig. 4-b) obtained in a different area but from the same patient. Normal acinar cells (cells responsible for digestive enzymes secretion) are distinctively grouped in acini clusters (black circles) and surrounded by thin collagen fibres (red arrows). The nuclei distribution is also homogeneous (one nucleus per acinar cell (grey arrows)). Larger regions are shown in supplementary figure S6. Pancreatic

islets (islets of Langerhans) which secrete hormones, could also be evidenced using SRS and gave a strong signal compared to the surrounding pancreatic tissue (data not shown).

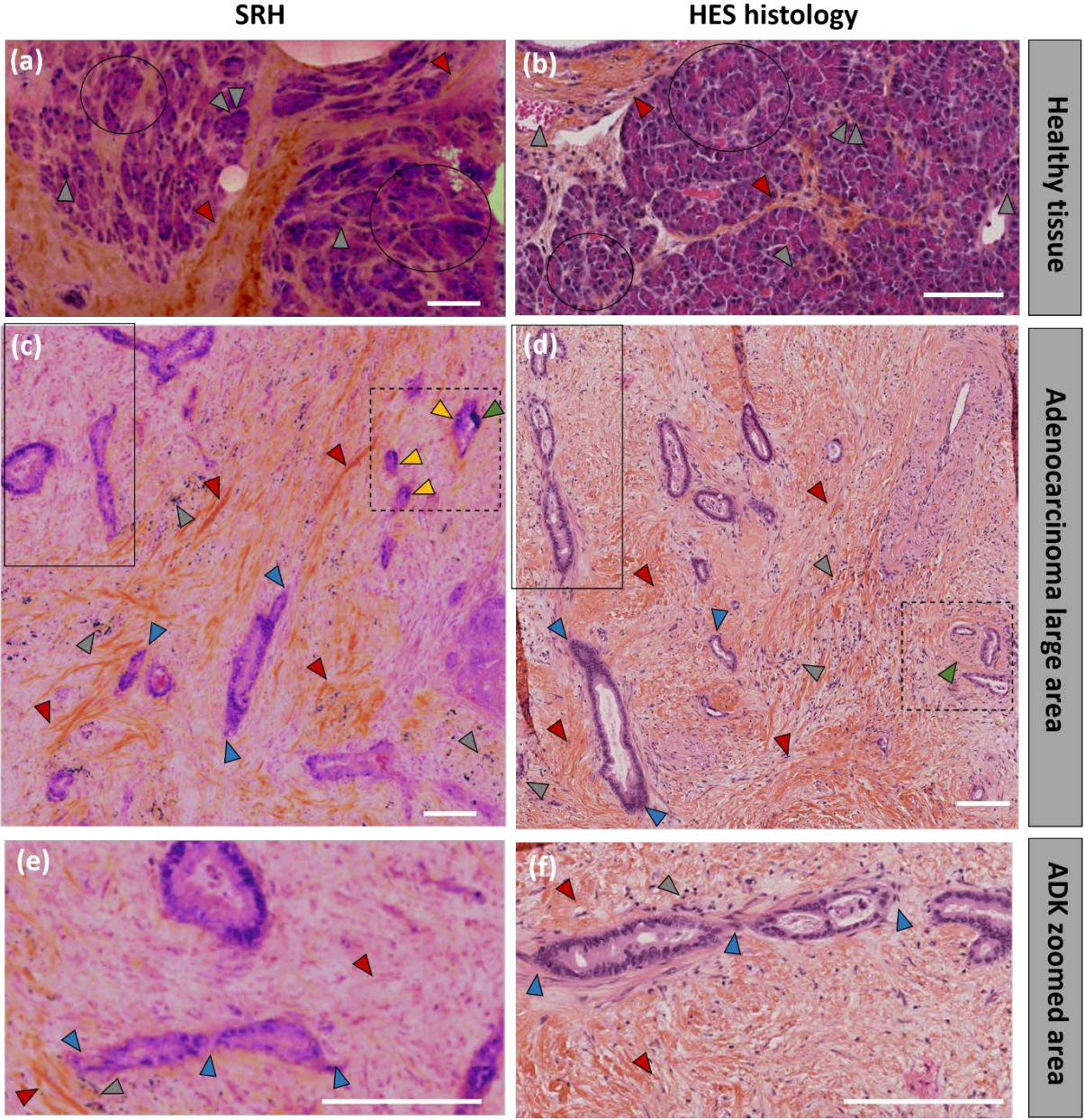

**Figure 4:** Adenocarcinoma diagnosis on human pancreas using Stimulated Raman Histology (SRH). (a) SRH picture of healthy (ie non-cancerous) pancreas tissue. (b) HES staining of the same of the region (a few mm$^3$) of the same patient. In both images acinar cells in group (black circles) surrounded with thin collagen fibers (red arrows) can be visualized. (c) SRH picture of a millimeter-scale region of a cancerous pancreatic tissue, (d) HES picture of a larger zone of the same patient in the same region (a few mm$^3$). In both cases, strong stroma fibrosis (red arrows), large angulated glands (blue arrows) can be observed. In the SRH image, a gland demonstrating both a nucleolus (yellow arrow) and an hyperchromatic cell (green arrow) in the same gland allowed to diagnose for adenocarcinoma while a nerve 'encapsulating' 2 glands in the HES picture (green arrow)

lead to the same diagnostic. (e) and (f) are zoomed on the solid line ROIs defined in (c) and (d) respectively in wich the same features are highlighted.

Figure 4 e and f depicts tumorous regions of a pancreatic tissue. This is the situation where a strong stroma fibrosis (red arrows) develops as a reaction of the host cells to the tumour focus[16]. The fibrosis modifies the surrounding micro-environment of the tumour and participates to its biological development mechanically and by promoting angiogenesis[17]. In striking contrast with the healthy case (Fig. 4a-b) the tissue disorganization affect not solely the stroma but also the acinar gland cell clusters, the persisting glands (blue arrows) are very disparate both in shape and size and spatially sparse within the connective tissue network. In that case the pancreas will undergo self-necrosis, which may be what the darker nuclei within the connective tissue on fig 4 c reveal (grey arrows). A closer look at the gland structures and surrounding micro-environment is presented on the magnified ROI from figures 4-c and d (Fig. 4 e-f) and allows to visualize further the described features.

Pancreatic diagnoses are usually delicate to make as the morphological changes induced by a benign inflammation (pancreatitis) and an adenocarcinoma (malignant tumour) are quite similar while the consequences and treatments will be drastically different for the patient. To differentiate between the two, clinicians search for abnormalities within the tissue. These are highlighted with dashed boxes in Fig. 4 c and d. In the SRH image both a hyperchromatic nucleus (green arrow) and nucleoli (yellow arrow) are present within the same gland (two nuclei irregularities co-exist within the same gland) which is a typical anomaly for adenocarcinoma. Other nucleoli found in the small glands in the close surrounding tissue (yellow arrows) fortify the adenocarcinoma diagnosis. In the HES image a nerve is wound around several glands (green arrow) and leads to the same adenocarcinoma diagnosis (fig 4 d). All together these results establish further the aptitude of SRH for identifying pancreatic cancer development.

*Fast SRH imaging using frequency modulated SRS (FM-SRS)*

The previous images of GI tissues have been performed using the $\lambda$-switch modality where a 1 mm x 1 mm stitched image required approximately 50 min acquisition time. In order to be useful in an intra-operative context and help in diagnostic making during surgeries, SRH should be as fast as extemporaneous histology (EH) that can screen a 1 mm x 1 mm sample in approximately 30 minutes. As already mentioned EH protocol uses toluidine blue only and provides less information than a full HES staining. Providing HES imaging in the time scale of EH would provide an unprecedented tool to delineate tumor margins with evident clinical benefits. To achieve this goal we have designed a new imaging platform (Fig. 1-a and S1) enabling the simultaneous acquisition, at the pixel level, of all the critical signals enabling instantaneous SRH (Fig. 5 h-g). As described in the materials and methods section the developed FM-SRS modality provides simultaneously the nuclei ($SRS_{nuclei} = SRS_{2930cm-1} - SRS_{2845cm-1}$) (Fig. 5-a), cell bodies (CARS – Fig. 5-b or TPEF – Fig. 5-c) and collagen (SHG – Fig. 5 d) images. Fig 5-h displays a 1 mm x 1 mm FM-SRS image of a healthy colon cryosection. This image is similar to figures 2-a and S3-a where crypts are homogeneous in size and shape, vacuoles are uniformly present within the glands and nuclei are both in the basal position and regularly spaced. Moreover the SHG image (Fig. 5 d) clearly highlights the thin collagen network (muscularis propria) circularly surrounding the Lieberkühn glands (Fig 5 f-g). At the bottom part of figure 5-h, glands appear longer and many nuclei are present. Because no muco-secretion loss is noticeable and since the glands are oval this does not indicate malignant morphological changes but manifests a change of the gland orientations due to the villi architecture (longitudinal sectioning instead of transverse). FM-SRS presents the advantage of reducing the acquisition time by a factor of 2 and the image shown in Fig. 5-h was achieved in 25 minutes (instead of 50 min with the $\lambda$-switch modality). In addition, FM-SRS can handle fast sample movements since all signals are acquired at the single pixel level, this is important to image fresh tissue sections that are prone to deformation during the image acquisition process (Fig. S7).

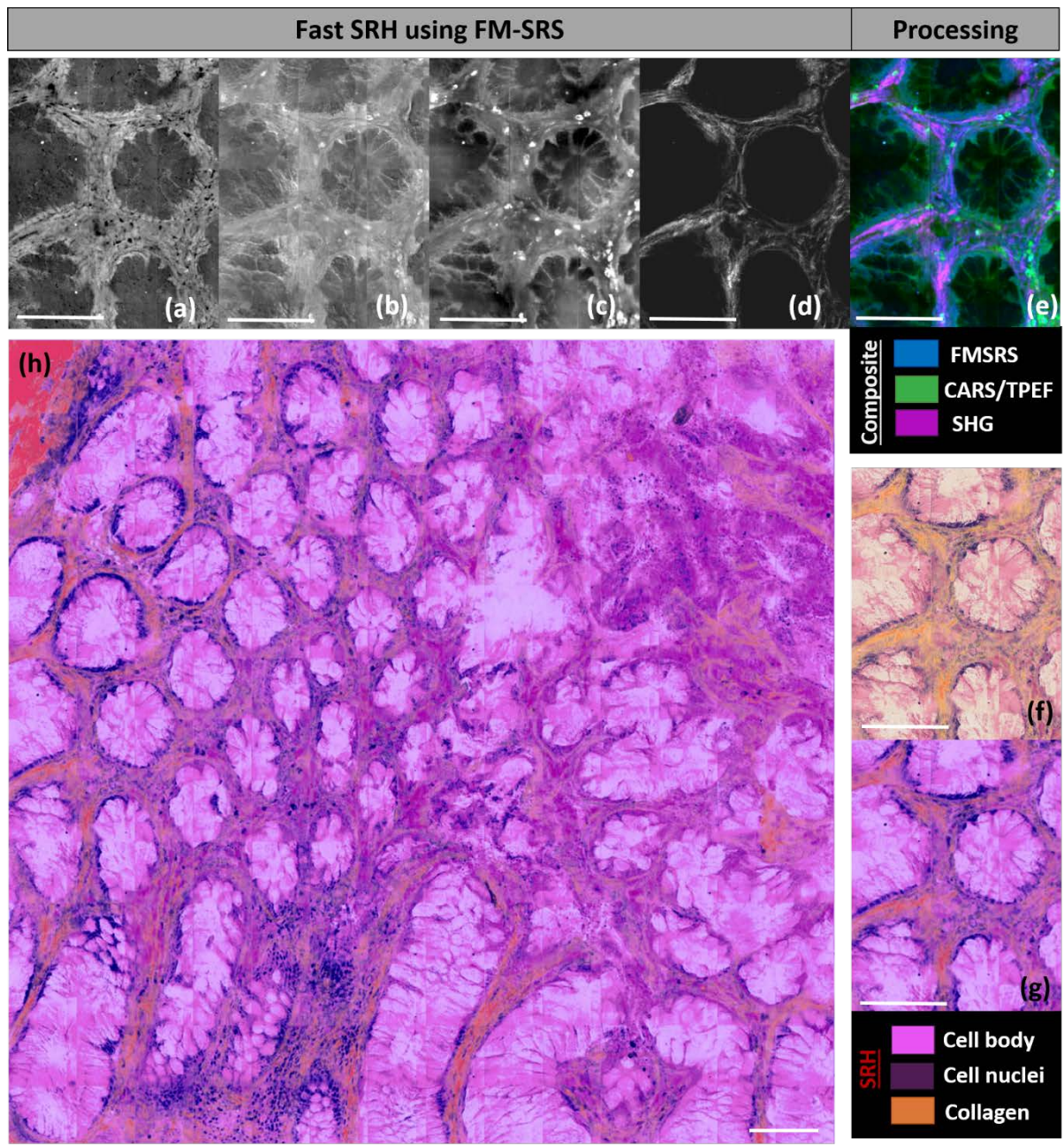

**Figure 5:** Fast SRH using the FM-SRS modality. (a) FM-SRS ($SRS_{nuclei}= SRS_{2930cm-1} – SRS_{2845cm-1}$): nuclei distribution, (b) CARS, and (c) TPEF: cell body distribution and (d) SHG: collagen distribution are acquired simultaneously. Combining (a-d), a composite image can be built: FMSRS (blue), TPEF (green) and SHG (magenta). Applying the LUTs described in the Materials and Methods section a SRH image can be built (f-g-h). (f) SRH from raw data, (g) image smoothed and color adjusted from (f). (h) SRH on a large millimeter-scale region on healthy (i.e. non-cancerous) human colon tissue. Acquisition time 25 minutes.

**Discussion**

Fast, reliable and accurate intraoperative diagnostic are essential during GI tumour surgery. Standard techniques such as extemporaneous histology (EH) are prone to artefacts as compared to standard HES protocols. However the 24-72h preparation time for HES slides precludes its implementation in an intra-operative context. Providing a reliable imaging framework for intraoperative histology with HES quality would ensure more efficient diagnosis and provide guidance for the surgery. In this context the near-perfect concordance of SRH images with standard HES, its compatibility with the operatory room time scale and workflow raises an invaluable opportunity. Here we reported the first demonstration of SRH imaging in GI tract cancer development and show how it can be used to create images with near-perfect accordance with standard HES protocol. We exemplified the performances of SRH images on both colon and pancreas normal and cancerous tissue sections. In all cases pathologist could perform a diagnosis based on SRH images, with an accuracy similar to standard HES. The developed imaging platform can detect the cell nuclei a key features for cancer diagnostic, this is made possible by the use of SRS that can quantify the ratio between $CH_3$ and $CH_2$ with a resolution of less than a micron. Because this ratio is also affected in collagen, our framework uses SHG to unambiguously identify collagen. This is important in the case of pancreatic tumours which exhibit abundant connective tissues. The generation of HES like images from SRS and SHG images is obtained with a dedicated virtual colouring process (see Materials and Methods section and supplementary Fig. S2). There is clearly room for improvement here using automated treatment based on HES nuclei, cell body and collagen colour identification and its transfer to SRS and SHG images.

In order to speed up the SRH acquisition time we have also demonstrated a novel imaging platform, FM-SRS, able to perform SRH images of GI tissues over millimetres large areas sections in a time that is compatible with the operatory room workflow. Most importantly our fast FM-SRS modality is insensitive to tissue movement during the image acquisition process because all the relevant signals are acquired simultaneously. This is important to image excised thick fresh tissues that are prone to deform. In order to test the ability of our FM-SRS imaging platform to image fresh tissue supplementary figure S7 presents a composite and SRH image obtained from a 1mm thick fresh colon tissue directly squeezed between two cover slips. The key features of healthy colon are clearly identifiable such as the homogeneous gland with vacuoles uniformly present within the glands. Such results open the route towards intra-operative SRH imaging to detect metastasis lymph nodes in the peritoneal cavity during gastric surgeries, such detection in the operatory theater would have an immediate impact on the patient survival rate.

**Conclusion**

In this paper, stimulated Raman scattering (SRS) microscopy was used in association with second harmonic generation (SHG) microscopy to visualize tissue sections from healthy and cancerous human gastro-intestinal tract. We demonstrated that this combination allows to perform stimulated Raman histology (SRH), a label-free technique providing histology images by assigning virtual colours to mimic haematoxylin, eosin and saffron (HES) staining. Different colon and pancreas cancers from patients could successfully be diagnosed using SRH images that show quasi-perfect concordance with standard HES staining. Both the general architecture alterations of the tissue and morphological changes at the subcellular level could be investigated by SRH microscopy on millimetre size GI tract tissues. Being compatible with the operatory room workflow SRH may ultimately be used as an additional extemporaneous histology tool to help in GI surgery decision making. Further clinical studies will be necessary to validate these results and identify the clinical cases that could benefit from this technology. Ultimately SRH technology could be miniaturized to millimetre size probes[18] to extend its ability to perform instantaneous histology in vivo.

**Conflict of interest**

The authors declare no conflict of interest

**Acknowledgements:**

We acknowledge financial support from the Centre National de la Recherche Scientifique (CNRS), Aix-Marseille University A*Midex (ANR-11-IDEX-0001-02) (A-M-AAP-ID-17-13-170228-15.22-RIGNEAULT), ANR grants France Bio Imaging (ANR-10-INSB-04-01) and France Life Imaging (ANR-11-INSB-0006) infrastructure networks and Plan cancer INSERM PC201508 and 18CP128-00.

**Contribution:**

B.S., R.C., X.A. performed the experiments; B.S., E.S., J.W. processed the data; B.S, F.P, F.C, M.G, H.R. analysed the data; H.R. and M.G. designed the research, all authors wrote the manuscript